\documentclass{aa}

\usepackage{natbib}
\usepackage{subfigure}
\usepackage{graphicx}
\usepackage{float}
\usepackage{natbib}
\usepackage{dcolumn}
\usepackage{bm}
\usepackage{ulem}
\usepackage{color}
\usepackage{txfonts}
\usepackage{rotating}
\usepackage{booktabs}
\usepackage{array}
\usepackage{tabularx}
\usepackage{threeparttable}
\usepackage{diagbox}
\usepackage{multirow}
\usepackage{makecell}
\usepackage{longtable}

\begin{document}

\title{Hydrogen-free Wolf-Rayet stars: Helium stars with envelope-inflation structure and rotation}


\author{Xizhen Lu\inst{1,2}
\and    Chunhua Zhu\inst{1*}
\and    Helei Liu\inst{1}
\and    Sufen Guo\inst{1}
\and    Jinlong Yu\inst{3}
\and    Guoliang L\"{u}\inst{2,1*}
}
 \institute{School of Physical Science and Technology, Xinjiang University, Urumqi, 830046, China\\
              \email{chunhuazhu@sina.cn }
         \and Xinjiang Astronomical Observatory, Chinese Academy of Sciences, 150 Science 1-Street, Urumqi, Xinjiang 830011, China\\
              \email{guolianglv@xao.ac.cn }
         \and College of Mechanical and Electronic Engineering, Tarim University. 843300, Alar, P.R.China   }



  \abstract
   {Observations have shown that the effective temperature of hydrogen-free Wolf-Rayet (WR) stars is considerably lower than that of the standard model, which means that the radius of the observed H-free WR stars is several times larger than that estimated by the standard model. The envelope inflation structure (EIS) caused by the radiation luminosity being close to the Eddington luminosity in the iron opacity peak region of H-free WR stars may be the key to resolve the radius problem of H-free WR stars.}
   {We study the structure and evolution of helium (He) stars with the EIS and discuss the influence of rotation on these He stars. We try to explain the H-free WR stars observed in the Milk Way (MW) and the Large Magellanic Cloud (LMC) by the He stars.
   We aim to explain the radius problem of H-free WR stars observed in the MW and the LMC through the He stars.}
   {Using the Modules for Experiments in Stellar Astrophysics code, we compute the evolution of He stars with and without MLT++ prescriptions and discuss their effects on the EIS. We have calculated the evolution of He stars using a new mass-loss rate formula and three different relative rotational velocity and compared our results with observations on Hertzsprung-Russell diagrams.}
   {The EIS has different effects on the structure and evolution of He stars with different masses. Due to the luminosity well below the Eddington limit, low-mass He stars with an initial mass of less than 12 $M_\odot$ do not produce EIS with or without the MLT++ prescription. High-mass He stars with an initial mass exceeding 12 $ M_\odot$ and without the MLT++ prescription produce the EIS. Since EIS is Eddington factor $\Gamma$-dependent, its radius increases with the increase of metallicity and decreases with rotational velocity increase. For rotating low-mass He stars, since the rotational mixing timescale is smaller than the evolutionary timescale, rotational mixing can increase the lifetime and allow He stars to evolve into WC stars during the helium giant phase. For rotating high-mass He stars, since rotation increases the mass-loss rate, the radius of the EIS decreases as rotational velocity increases. The rotation-decay timescale of rapidly rotating He stars is very short, and the rapidly rotating He stars only appear within the first one tenth of their lifetime, which is consistent with the observations of WR stars.}
   {The low luminosity (log$(L/L_{{\odot}})\leq5.2$) H-free WR stars in the MW and the LMC can be explained by the helium giant phase in low-mass He stars, the high $X_{C}$ and $X_{O}$ in WC stars can only evolve through low-mass He stars with a rapid rotation. High-mass He stars with the EIS can explain H-free WR stars with a luminosity exceeding $10^{5.7} L_{{\odot}}$ and an effective temperature above $10^{4.7}$ K in the MW. They can also explain H-free WR stars on the right-hand side of the He zero-age main sequence in the LMC. High-mass stars with the EIS evolve into WO stars at the final evolution stage, and the shorter lifetime fraction is consistent with the small number of observed WO stars.}


   \maketitle
%
\section{Introduction}
Wolf-Rayet (WR) stars are generally helium burning, evolved massive stars that are characterized by intense and broad emission lines in the spectrum \citep{Beals1940, Chiosi1986, Maeder1994}.
WR stars have lost most of their hydrogen envelope via stellar wind or mass transfer through the Roche-lobe overflow in close binary systems \citep{Conti1975,Crowther2007}.
Compared with other normal stars, WR stars have high effective temperatures (log($T_{\rm eff}$/K)$\geq4.5$) and high luminosity (log$(L/L_{{\odot}})\geq4.8$).
Due to the strong radiation pressure on their surface, WR stars have powerful stellar winds with a typical mass-loss rate of $\dot{M}\gtrsim10^{-5} M_{{\odot}}\mathrm{yr}^{-1}$ \citep{Tramper2016,Hamann2019,Sander2019}.
Based on the presence of certain emission lines in their optical spectrum (i.e., depending on whether their optical spectrum is dominated by nitrogen, carbon, or oxygen lines), WR stars are broadly classified into three subtypes: nitrogen-type (WN), carbon-type (WC), and oxygen-type (WO) stars. WR stars can be classified as H-rich WN stars and H-free WN stars depending on their surface hydrogen abundance \citep{Hamann1995}.
All WC stars and WO stars have almost no hydrogen in their envelope\citep{Sander2012}.
H-free WR stars(i.e., H-free WN, WC, and WO stars) are the later evolution stage of WR stars.

The luminosity of the H-free WR stars is usually close to their Eddington limit, and energy in the envelope is mainly transported by radiation \citep{Petrovic2006,Grafener2012}.
The convection velocity in the envelope can approach the sound velocity, but convection transmission is still ineffective relative to radiation transmission \citep{Grafener2012,Grassitelli2016}.
Since the envelope does not produce energy, the luminosity is constant, but the Eddington luminosity limit $L_{\mathrm{Edd}}=4 \pi c G M / \kappa$ varies with the opacity, and the Eddington factor $\Gamma=\kappa L_{\mathrm{rad}} /(4 \pi c G M)$ is the largest at the iron opacity peak \citep{Hillier1998,Sanyal2015}.
Due to the influence of the iron opacity peak at the temperature of $10^{5.2}$ K, a region with a radiation luminosity close to the Eddington luminosity appears in the envelope\citep{Petrovic2006}.
This region, which has an extremely low density, causes the envelope radius to expand; thus, this region is called the envelope inflation structure (EIS) \citep{Grafener2012,Ro2016}.
As the opacity of the temperature below the iron opacity peak is significantly reduced, a density inversion phenomenon occurs in the outer layer of the EIS\citep{Grafener2012,Sanyal2015}.
\cite{Paxton2013} and \cite{Ro2016} studied and confirmed the existence of this structure and showed that it has a strong instability under one-dimensional simulations.
The strong instability of this structure presents significant computational challenges for the corresponding stellar evolution code\citep{Paxton2013}. In order to continue simulating the evolution, the code must employ additional techniques, such as extra mixing, increased convection efficiency, or other numerical methods, to eliminate the structure\citep{Agrawal2022}.
Eliminating the EIS results in the generation of H-free WR stars with a small radius $\left(R < 3R_{\odot}\right)$ and high effective temperature$\left(T_{\mathrm{eff}} \gtrsim 10^{5.2} \mathrm{K}\right)$ \citep{Poniatowski2021}.
The radius of H-free WR stars obtained from theoretical models without the EIS is an order of magnitude smaller than the observed radius, this discrepancy is also known as the WR radius problem.

The EIS is primarily dominated by the factor $\Gamma=\kappa L_{\mathrm{rad}} /(4 \pi c G M)$, which contains two independent variables, namely $\kappa$ and $L/M$. $\kappa$ is the Rosseland mean opacity, which is a function of the temperature, density, and chemical composition. The $L/M$ ratio is a function of the mean molecular weight of the He-burning core. In addition to $\Gamma$, another factor affecting the EIS is the mass-loss rate.
\cite{Petrovic2006} and \cite{Ro2016} studied the launching and structure of the radiation-driven wind of WR stars and found that a too strong radiation-driven wind can destroy the EIS.
The investigation of \cite{Grassitelli2018} also showed that the value of the mass-loss rate affects the effectiveness of the EIS.
\cite{Grafener2012}, \cite{Grassitelli2018}, and \cite{Ro2019} conducted detailed studies on the influence of the metallicity and the mass-loss rate on the EIS of the He zero-age main sequence(HeZAMS). However, it is well known that the mean molecular weight of the core increases with the stellar evolution , which can have an impact on the $L/M$ ratio. Therefore, the results of the HeZAMS do not apply to the evolved He stars.
The latest observations from Gaia suggest that previous studies may have overestimated the mass-loss rate of high-luminosity H-free WR stars \citep{Hamann2019,Sander2019}. Therefore, this work studies for the first time the evolution of H-free WR stars with the EIS using the updated mass-loss rate.

It is well known that rotation has an important effect on the evolution of massive stars \citep{Maeder2000,Aerts2004,Meynet2005,Maeder2010,Brott2011,cui2018,Zhao2020}.
Observations show a bimodal distribution of rotational velocities for early B-type and O-type stars, with a large proportion of stars having a rapid rotation ($v >200 \mathrm{~km} / \mathrm{s}$), with extremes of up to $500-600 \mathrm{~km} / \mathrm{s}$ \citep{deMink2013}.
A large fraction of the observed Wolf-Rayet (WR) stars
A significant proportion of the observed WR stars show line effects in their spectropolarimetry, suggesting the possibility of rapid rotation \citep{Grafener2012b,Graefener2013,Vink2017b,Abdellaoui2022}. Furthermore, WR stars characterized by a rapid rotation are also thought to be associated with long-duration gamma-ray bursts \citep{Yoon2006,Eldridge2006}.
The rotation causes additional instabilities, such as dynamical shear instability, secular shear instability, Eddington-Sweet circulation, and Goldreich-Schubert-Fricke instability \citep{Spiegel1970,Zahn1974,Heger2000,Wang2018}. These instabilities enhance angular momentum transport and chemical mixing \citep{Spiegel1970}.
The rotation-enhanced mixing changes the chemical composition of the envelope and core and then affects the $\kappa$ of the envelope and the $L/M$ ratio of the core.
The rotation also causes centrifugal forces that increase the mass-loss rate \citep{Maeder2010}. Therefore this work also studies the effect of rotation on the evolution of the H-free WR stars with EIS.

In this work, we study the effects of the EIS, stellar wind, and rapid rotation on the evolution of H-free WR stars. We compare our He-star models with the observations in the Hertzsprung-Russell (HR) diagram. In section 2, we introduce our model parameters.
In section 3, we explore the effects of the EIS on the evolution of stars and the  influence of the rotation and stellar wind within them. The conclusion is provided in section 4.
\section{Model}
H-free WR stars are essentially He stars with He burning at their core, so we can create He stars and start their evolution at the HeZAMS to avoid the uncertainty of the pre-H-free WR stage.
We employ the open-source stellar evolution code Modules for Experiments in Stellar Astrophysics(MESA, version15140 \citep{Paxton2011, Paxton2013,Paxton2015,Paxton2018}) to evolve He-star models with initial masses ranging from 4 to 35 $M_{\odot}$. To investigate the evolution of He stars, we have constructed a grid of H-free stars.
The models relax to chemical homogeneity, thermal equilibrium, and core temperature before reaching the He ignition.
The models evolve from the HeZAMS until all carbon is burned.
The main parameters in our models are the metallicity, convection, EIS, and mass-loss rate.

\subsection{Metallicity and convection}
The H-free WR stars observed are mainly concentrated in the Milky Way (MW) and the Large Magellanic Cloud (LMC).
We take the metallicity values for each galaxy. The initial metallicities ($\mathrm{Z}$) used in our models are: $\mathrm{Z}$=0.02 for the MW and $\mathrm{Z}$=0.008 for the LMC. The initial abundance of He is set to $\mathrm{Y}=1-\mathrm{Z}$.
The convection is calculated using the Ledoux criterion, and the mixing-length parameter $\alpha_{\mathrm{LMT}}$ = 1.5 is adopted \citep{cui2018,Lu2017}. The efficiency parameter of the semi-convection is set to $\alpha_{\mathrm{SEM}}$ = 1.0 \citep{Zhu2017,Lu2020}. We use the same step-function overshooting model as \cite{Brott2011} to calculate the overshoot area, where the overshooting parameter is 0.335.
\subsection{Envelope inflation structure}
In MESA codes, simulating the EIS and density inversions require extremely short time steps\citep{Paxton2013}.
Therefore, \cite{Paxton2013} proposed the MLT++ prescription, that is,
the stellar models require an additional convection efficiency, which is substantially larger than that permitted by the upper-limit estimate of the convection luminosity.
However, this results in the elimination of the EIS and density inversions.
In this work, we calculate the structural evolution of H-free WR stars with the EIS using the standard mixture length theory prescription; additionally, we also calculate the evolution of H-free WR stars without the EIS using the MLT++ prescription for control.
\begin{figure*}[htb]
\centering
\includegraphics[width=\textwidth]{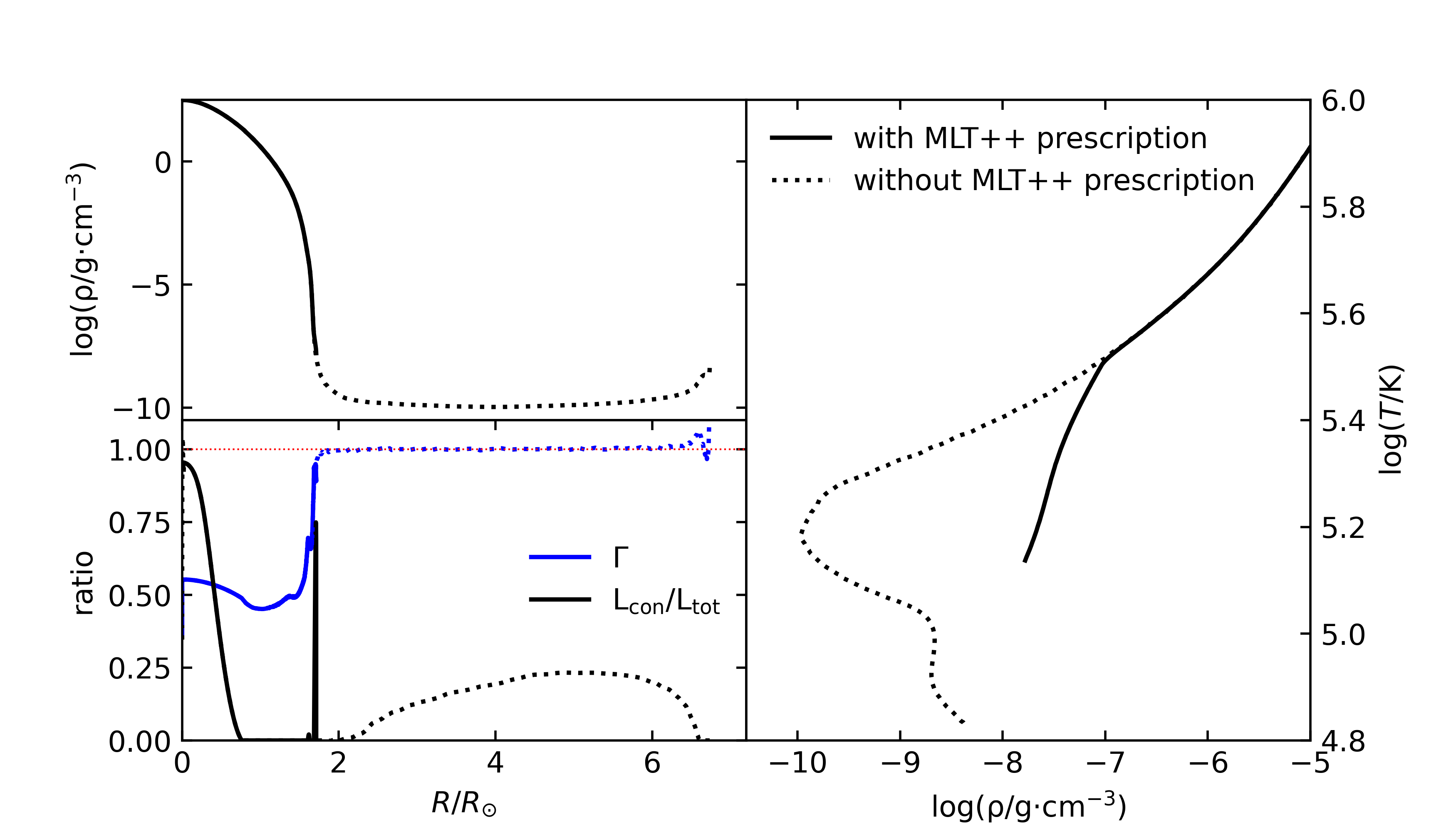}
\caption{Internal structure of a HeZAMS with a mass of $30 M_{\odot}$ at solar metallicity with (solid line) and without (dotted line) the MLT++ prescription. The top left panel shows density as a function of the radius, and the bottom left panel shows the luminosity ratio as a function of the radius. The right panel shows the relationship between temperature and density in the stellar envelope.}
\label{fig:inflation}
\end{figure*}
In order to show the effects of the EIS,
we calculate the structure of the HeZAMS $30 M_{\odot}$ model with and without the MLT++ prescription,
which is illustrated in Figure \ref{fig:inflation}. The envelope inflation greatly affects the structure around the surface of the He stars, and it can be found that the surface structure of He stars with and without MLT++ prescription is completely different.
The density-temperature curve (the right panel of Figure \ref{fig:inflation} ), shows that
there is a density inversion in the model without the MLT++ prescription.
Usually, the density is the lowest on the stellar surface, as shown by the solid line in Figure \ref{fig:inflation}.
However, the density at $T=10^{5.2}$ K in the model without the MLT++ prescription is the lowest,
and it is two orders of magnitude lower than that on the stellar surface,
which is triggered by the iron opacity peak. This results in stellar envelope inflation producing EIS.
The density of the EIS is extremely low($\log\rho\sim 10^{-10} \mathrm{g} \cdot \mathrm{cm}^{-3}$), so the convective transmission is inefficient, and the energy transmission is dominated by radiation.
As shown by the radius-luminosity ratio curve (the bottom left panel of Figure \ref{fig:inflation}), the ratio of the convective luminosity to the total luminosity (${L}_{\text {con }}/{L}_{\text {tot }}$) in the model without the MLT++ prescription is always less than 0.25, while the Eddington factor $\Gamma$ is always 1.
The radius-density curve (the top left panel of Figure \ref{fig:inflation}), shows that the stellar radius of the 30 $M_\odot$ HeZAMS without the MLT++ prescription
can reach around 7 $R_\odot$, but it is only around 1.8 $R_\odot$ for the model with it. These findings indicate that the EIS changes considerably the profile around the stellar surface.
\subsection{Mass-loss rate and rotation}
The mass-loss of WR stars is due to the iron line-driven wind, i.e, photons scattered by the absorption lines of iron group elements at the iron opacity peak \citep{Castor1975,Vink2005,Vink2012}.
Recently, several works have used the new stellar atmospheres models to show that the mass-loss rates of WR stars are $Z_{\mathrm{Fe}}$-dependent and Eddington parameter $\Gamma_{\mathrm{e}}$-dependent, and the mass-loss rate decreases with increasing $X_{\mathrm{C}}$ and $X_{\mathrm{O                            }}$ on the surface\citep{Sander2020a,Sander2020b}. Based on the works of \cite{Sander2020a} and \cite{Sander2020b}, we adopt a new expression for the mass-loss rate($\dot{M}_{SV}$):
\begin{equation}
\begin{aligned}
\log \dot{M}_{SV}=a \cdot \log \left[-\log \left(1-\Gamma_{\mathrm{e}}\right)\right]-\log (2) \cdot\left(\frac{\Gamma_{\mathrm{e}, \mathrm{b}}}{\Gamma_{\mathrm{e}}}\right)^{c}
\\
+\log \dot{M}_{\mathrm{off}}-0.285 X_{\mathrm{C}}-0.165 X_{\mathrm{O}}
\label{eq:1}
\end{aligned}
\end{equation}
where $a=2.932$, $\Gamma_{\mathrm{e}, \mathrm{b}}=-0.324 \cdot \log \left(Z_{\mathrm{Fe}} / Z_{\mathrm{Fe}\odot}\right)+0.244$, $c=-0.44 \cdot \log \left(Z_{\mathrm{Fe}} / Z_{\mathrm{Fe}\odot}\right)+9.15$, and $\log \dot{M}_{\mathrm{off}}=0.23 \cdot \log \left(Z_{\mathrm{Fe}} / Z_{\mathrm{Fe}\odot}\right)-2.61$.

This mass-loss rate expression is only applicable to the classical WR stellar luminosity range because it is $\Gamma_{\mathrm{e}}$-dependent\citep{Sander2020a}. Therefore, for luminosities lower than that of classical WR stars, we use the star wind equation proposed by \cite{Vink2017}:

\begin{equation}
\log \dot{M}=-13.3+1.36 \log \left(L / L_{\odot}\right)+0.61 \log \left(Z_{\mathrm{Fe}} / Z_{\mathrm{Fe}\odot}\right)
\label{eq:2}
\end{equation}

However, Gaia's observations have shown that the distance to WR stars derived from previous observations is not well constrained\citep{Hamann2019,Sander2019}.
Some H-free WR stars have a higher luminosity than previous estimates, which means that the $\Gamma_{\mathrm{e}}$ factor in the mass-loss rate expression may be overestimated.
Furthermore, \cite{Poniatowski2021} showed that it is not sufficient to use a wind model involving the co-moving frame radiative transfer to simulate classical WR stars.
Due to the uncertainty in the mass-loss rate of high-mass He stars, we consider the \cite{Moens2022} 3D radiation-hydrodynamic simulations of WR winds, which suggest that the mass-loss rate formula calculated by \cite{Sander2020b} may be overestimated. Therefore, we introduced a wind factor ($f_{\mathrm{sv}}$) to the mass-loss rate equation, resulting in the formula: $\dot{M} = f_{\mathrm{sv}}\cdot\dot{M}_{\mathrm{sv}}$.

The angular momentum transfer and the element mixing due to the rotational instabilities should also be calculated in the model. These instabilities enhance the angular momentum transport and chemical mixing \citep{Spiegel1970}.
Following the results of \cite{Zhu2017}, \cite{Zhu2021}, and \cite{Wu2021}, the ratio of the
turbulent viscosity to the diffusion coefficient ($f_{\mathrm{c}}$) and the ratio of the sensitivity to the chemical mixing coefficient ($f_{\mu}$) are assumed to be 0.0228 and 0.1, respectively. He stars of different masses have different critical rotation velocities, so we use the ratio of the rotation angular velocity to the critical rotation angular velocity ($\omega/\omega_{\text {crit }}$) to represent a given rotation.

Rotation can also enhance the mass-loss rate. Centrifugal force from rapid rotation also increases mass-loss rates\citep{Langer1998}:
\begin{equation}
\dot{M}=\left(\frac{1}{1-\Omega / \Omega_{\text {crit }}}\right)^{\beta} \dot{M}_{v_{\text {rot }}=0}
\label{eq:3}
\end{equation}
where $\Omega$ and $\Omega_{\text {crit }}$ are the angular velocity and the critical angular velocity, respectively, and $\beta=0.43$ \citep{Langer1998,Zhu2017,Wu2021}.

Considering that rotation gives rise to deformation and gravity darkening, there are other hypotheses regarding the influence of rotation on the mass-loss rate. For example, \cite{Muller2014} studied the two-dimensional rotation model with gravity darkening and showed that the rotation reduces the mass-loss rate. However, it is difficult for the one-dimensional code MESA to use the mass-loss rate based on the two-dimensional model.
In addition, we reduce the wind factor to decrease the mass-loss rate, and this can be regarded as an approximate effect in which rotation can reduce the mass-loss rate.

\section{Results}
\begin{figure}[htb]
\centering
\includegraphics[width=0.5\textwidth]{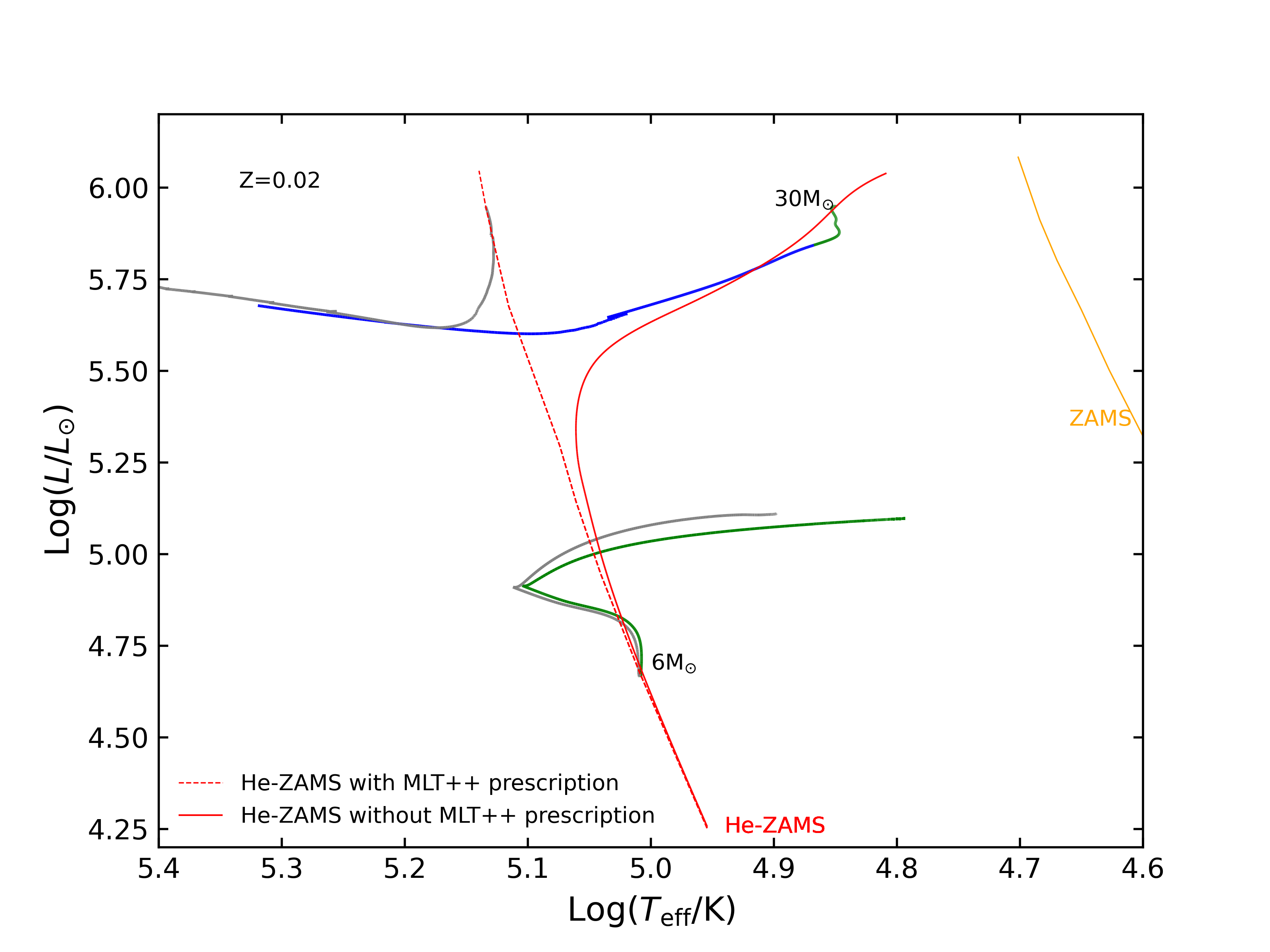}
\caption{HR diagram of the HeZAMS with and without the EIS and evolutionary tracks of He stars with 6 and 30 $M_{\odot}$.
The green and blue lines represent the WN and WC sequences, respectively. The gray lines represent the evolutionary track of the He stars without the EIS, obtained using MLT++. The red dotted line represents the HeZAMS without the EIS, while the red solid line  represents the HeZAMS with the EIS. The orange line corresponds to the ZAMS of normal stars.}
\label{fig:hl}
\end{figure}
Following \cite{Georgy2012}, we consider He stars with a surface carbon abundance less than the nitrogen abundance ($\frac{X_\mathrm{{C}}}{X_\mathrm{{N}}}<1$) as H-free WN stars. Otherwise, they are considered as WC/WO stars.
In the present paper, we simulate the evolution of He stars with and without the EIS with masses of 4, 6, 8, 10, 12, 16, 20, 25, 30 and 35 $M_{\odot}$.

\subsection{Evolution of helium stars}
To show the effect of the EIS on the evolution of He stars with different masses, we calculate the position of the HeZAMS with and without the MLT++ prescription, as well as the evolutionary tracks of He stars with initial masses of 6 and 30 $M_{\odot}$, as shown in Figure \ref{fig:hl}.
The EIS is $\Gamma$-dependent, so it depends strongly on the stellar luminosity.
Compared with the HeZAMS with MLT++ prescription, the HeZAMS without MLT++ prescription forms EIS and shifts toward a lower temperature at a high luminosity.
This result is consistent with those of \cite{Petrovic2006}, \cite{Grafener2012}, and \cite{Ro2016}.
Then, in this work, we divide He stars into two types: low-mass He stars without a significant EIS that have an initial mass lower than 12 $M_{\odot}$ and, high-mass He stars with a significant EIS that have an initial mass higher than 12 $M_{\odot}$.
In order to show the different evolutionary tracks, we select two typical models: the 6 $M_{\odot}$ model for low-mass He stars and the 30 $M_{\odot}$ model for high-mass He stars.

\subsubsection{Low-mass helium stars}
As shown for the 6 $M_\odot$ He star in Figure \ref{fig:hl}, the evolutionary track of He stars with the MLT++ prescription is almost the same as that of He stars without the MLT++ prescription. The main reason is that the luminosity of low-mass He stars is significantly lower than their Eddington luminosity, and they cannot produce the EIS with or without the MLT++ prescription.
During the He main sequence, the rate of nuclear energy generation per unit mass $\epsilon_{\text {nuc }}$ increases, and the He star contracts, so the luminosity and effective temperature increase with the star evolution.
During the helium giant stage, the He stars evolve into a helium giant structure: a high-density core with a low-density envelope expanded by the He-shell burning. The luminosity produced by the He-shell burning is larger than the luminosity that can be transported by the envelope, so the excess energy is absorbed by the envelope. The internal energy in the envelope increases, and then the He star expands, so the star radius increases, and the effective temperature decreases.
He stars in the helium giant stage always show the characteristics of H-free WN stars because they always have a large He-rich envelope remaining.

\subsubsection{High-mass helium stars}
\begin{figure}[htb]
\centering
\includegraphics[width=0.5\textwidth]{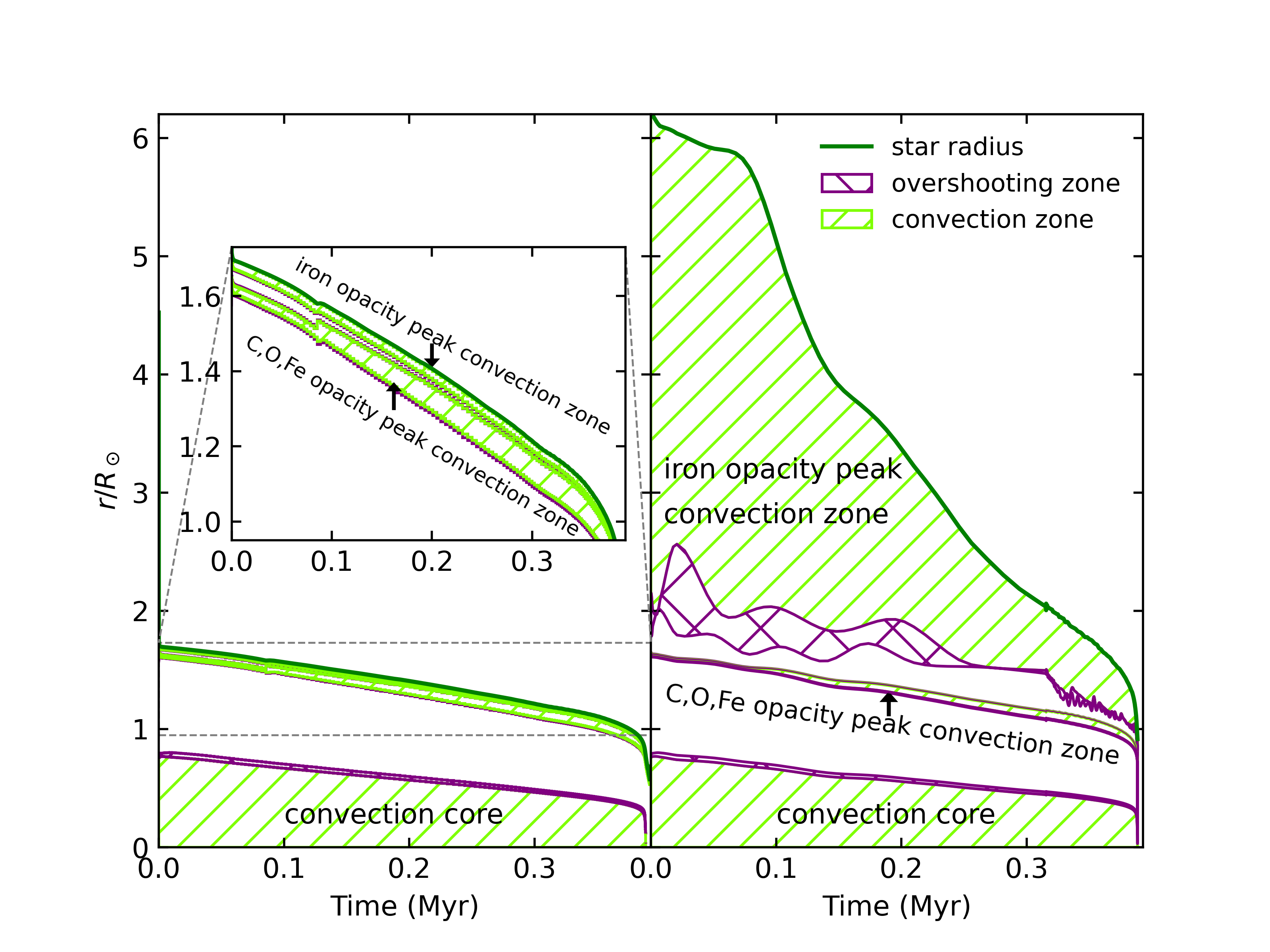}
\caption{Structure evolution (Kippenhahn) diagrams. The x-axis shows evolutionary time(unit: million years; zero point: HeZAMS), and the y-axis represents the radius. The left and right panels show the 30 $M_{\odot}$ He star with and without MLT++ prescription, respectively.
The green-shaded and purple-shaded regions indicate the convection zone and overshooting zone, respectively.}
\label{fig:kipp}
\end{figure}
As shown for the 30 $M_\odot$ He star in Figure \ref{fig:hl}, compared with the evolutionary track of He star without the MLT++ prescription, the evolutionary track of the He star without the MLT++ prescription is clearly shifted toward a lower temperature.
Since their radiative luminosity in the iron opacity peak convection zone(FeCZ) is close to the Eddington limit. The high-mass He stars without the MLT++ prescription have an EIS, so that the evolution tracks of high-mass He stars can start from a lower effective temperature.
The He star with MLT++ prescription increases the convection efficiency to reduce the Eddington factor $\Gamma$, so the EIS is eliminated and the envelope becomes more dense.
As shown in Figure \ref{fig:kipp}, He stars with the MLT++ prescription in which the radius of the FeCZ is always only a few percent of the solar radius.
For He stars without the MLT++ prescription, the strong radiation pressure makes the density of the FeCZ to be extremely low, so the radius of the FeCZ increases by two orders of magnitude, i.e., EIS.
During the early stage of the He main sequence, the mass of He star is rapidly reduced due to the high mass-loss rate, and their luminosity decreases.
Furthermore, with the He-core burning, the average molecular weight ($\mu_{\mathrm{avg}}$) and $L/M$ of the core increase.
The radius of EIS decreases slowly at this stage under the combined effect of increasing $L/M$ and $\mu_{\mathrm{avg}}$.
In the later stage of the He main sequence, the He-rich envelope has been completely stripped by the high mass-loss rate. The CO-rich envelope is exposed, which has a higher $\mu_{\mathrm{avg}}$ and a higher convective luminosity than the He-rich envelope, so the radius of EIS decreases rapidly.
As a result, the radius of the He star decreases, and the effective temperature increases.
At this stage, the $X_{\mathrm{C}}$ and $X_{\mathrm{O}}$ on the stellar surface increase, and high-mass He stars can evolve into WC/WO stars.

\subsection{Effects of the input parameters}
The effects of the metallicity and wind factor ($f_{\mathrm{sv}}$) on the EIS of He stars were discussed by \cite{Petrovic2006} and \cite{Grassitelli2018}. Similar to their results, we find that a high metallicity can enhance the EIS, while a large $f_{\mathrm{sv}}$ can trigger a high mass-loss rate, which destroys the EIS. In this work, we focus on the effects of rotation.

\begin{figure}[htb]
\centering
\includegraphics[width=0.45\textwidth]{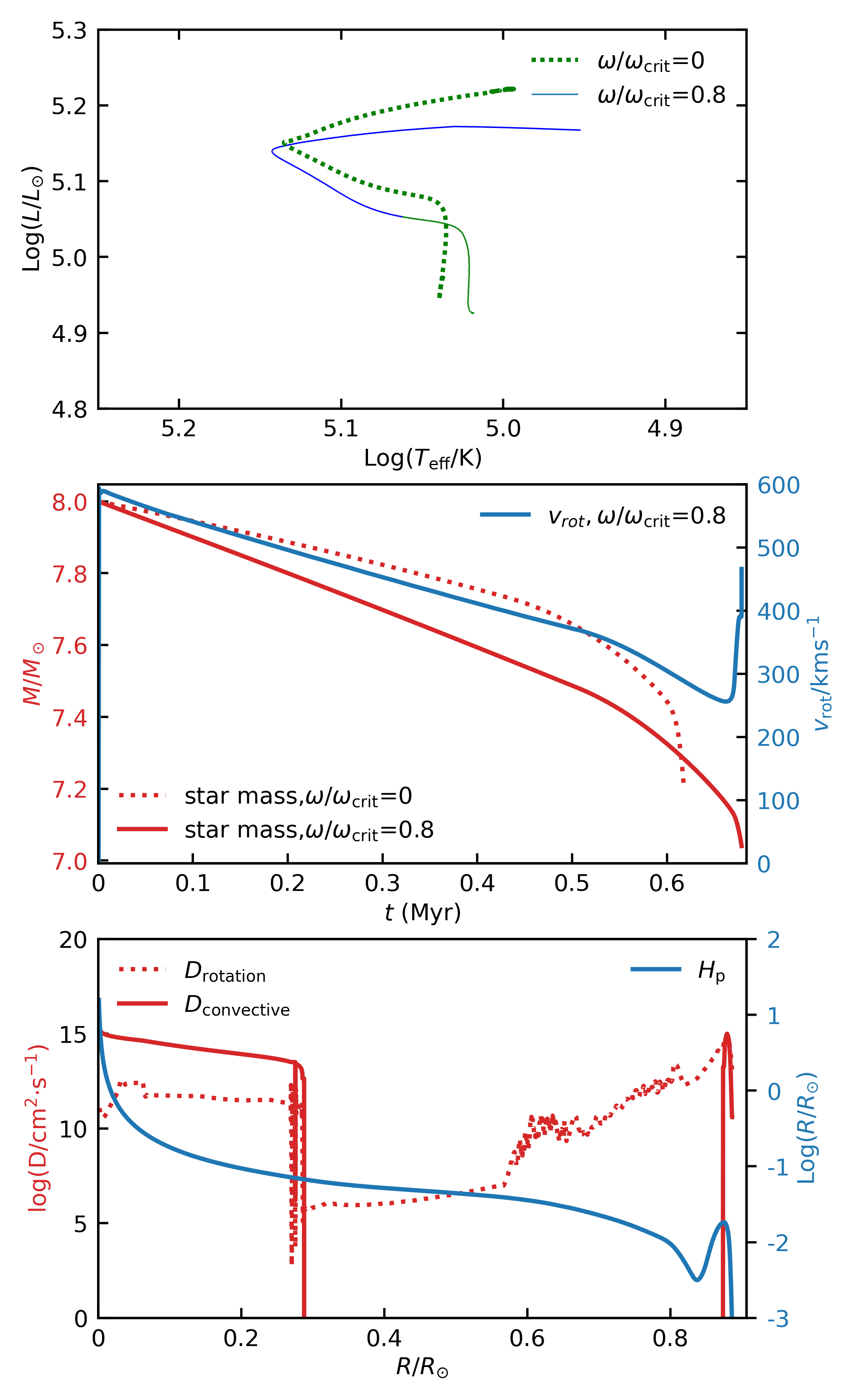}
\caption{The upper panel shows the HR diagram of the evolutionary tracks of 8 $M_{\odot}$ He stars with $\omega/\omega_{\text {crit }}$ =0 and $\omega/\omega_{\text {crit }}$ =0.8. The middle panel shows the evolution of rotation rate and mass with time. The x-axis shows time(unit: million years; zero point: HeZAMS). The red dotted and solid lines represent He stars with $\omega/\omega_{\text {crit }}$ =0 and $\omega/\omega_{\text {crit }}$ =0.8, respectively, corresponding to the red y-axis on the left. The blue solid line represents the rotational velocity of the He star with $\omega/\omega_{\text {crit }}$ =0.8, corresponding to the blue y-axis on the right. The lower panel shows the stellar mixing coefficients and the pressure scale height as a function of radius at HeZAMS. The red solid and dotted lines represent the convective mixing coefficient and the rotational mixing coefficient, respectively, corresponding to the red y-axis on the left. The blue solid line represents the pressure scale height, corresponding to the blue y-axis on the right.}
\label{fig:08r}
\end{figure}

Rotation mainly affects mixing and the mass-loss rate. For low-mass He stars, we take the model of an 8 $M_\odot$ He star as an example, as shown in the Figure \ref{fig:08r}.
From the HR diagram we can see that the rotation has little effect on the evolutionary tracks of the stars.
The centrifugal force causes He stars to expand at their latitudes, which results in a slight reduction in effective temperature and luminosity, consistent with the results for main sequence stars \citep{Brott2011}.
We also note that, unlike the non-rotating He stars that are always WN stars, the rotating He stars can evolves into the WC/WO stars. As shown in the middle panel of Figure \ref{fig:08r}, since we have a rotational centrifugal force enhancement term in the mass-loss rate formula, the rotation-enhanced mass-loss rate also causes the rotating He star to have an additional loss of only 0.15 $M_{\odot}$ over its entire lifetime.
Simultaneously, rapid rotation can trigger rotational mixing and increase the core mass, which results in a rotating He star having a longer lifetime than a non-rotating He star.
The low mass-loss rate also takes away less angular momentum, so the rotational velocity of the rotating He stars has always been above 250 $\mathrm{km}\cdot\mathrm{s}^{-1}$. The rotational mixing diffusion for an element $i$ is followed by the equation:
\begin{equation}
\rho \frac{\partial X_i}{\partial t}=  \frac{1}{r^2} \frac{\partial}{\partial r}\left[r^2 \rho D_{\mathrm{rot}} \frac{\partial X_i}{\partial r}\right]
\label{eq:4}
\end{equation}
where $D_{\mathrm{rot}}$ is the rotational mixing coefficient, $r$ is the radial coordinate, and $\rho$ is the local density.
For a region where the diffusion velocity $v_{i}$ of an element $i$ is relatively constant, the rotational mixing timescale follows the equation:
\begin{equation}
\tau \approx \frac{L}{v_{i}}=\frac{L p}{\rho g D_{\mathrm{rot}}}=\frac{L H_{\mathrm{p}}}{D_{\mathrm{rot}}}
\label{eq:5}
\end{equation}
where $L$ is the height of the region and $H_{\mathrm{p}}$ is the pressure scale height.
As shown in the lower panel of Figure \ref{fig:08r}, the radiation zones at 0.29 $R_{\odot}$ to 0.39 $R_{\odot}$ near the convection core with $H_{\mathrm{p}}\sim10^{-1.2} R_{\odot}$ and $D_{\mathrm{rot}}\sim10^{6.5}\mathrm{cm}^2 \mathrm{s}^{-1}$, and the rotational mixing timescale $\tau\sim3\times10^{5} \mathrm{yr}$. It is shorter than the lifetime ($\sim7\times10^{5} \mathrm{yr}$) of He star.
Thus the rotational mixing increases the He star lifetime and the increase in $X_{\mathrm{C}}$ and $X_{\mathrm{O}}$ on the surface makes He star evolve into WC/WO stars in the helium giant phase.

\begin{figure}[htb]
\centering
\includegraphics[width=0.45\textwidth]{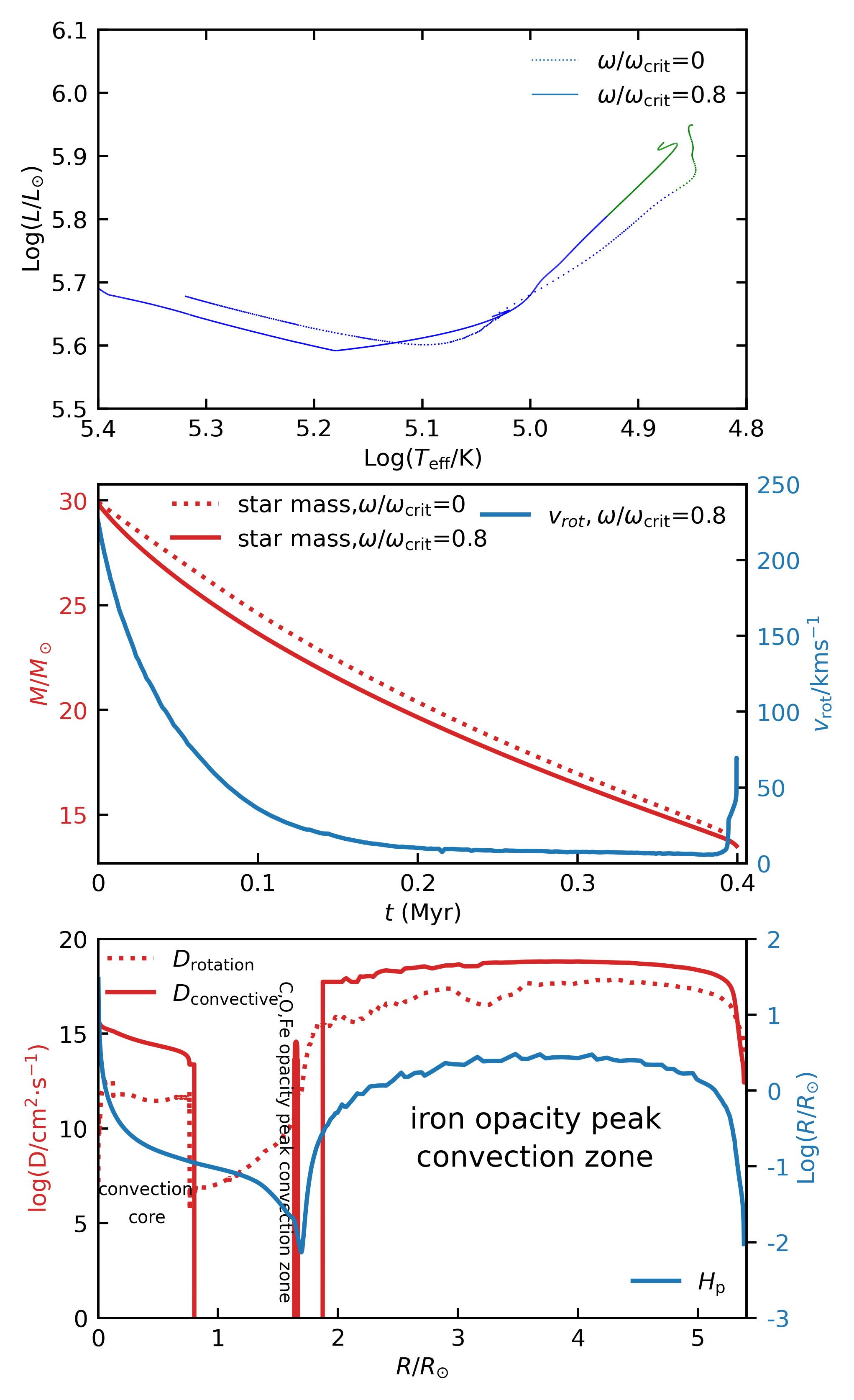}
\caption{Same as Fig.\ref{fig:08r}, but for the models with $M=30 M_{\odot}$}
\label{fig:02r}
\end{figure}
For high-mass He stars, the higher He star mass, the higher $\Gamma$ is. The He star models without MLT++ prescription have significantly larger radii by producing the EIS at high $\Gamma$.
Since the critical rotational velocity is $\Gamma$-anti-dependent and $R$-anti-dependent ($v_{\text {crit }}=\left(G M\left(1-\Gamma\right) / R\right)^{1 / 2}$), the critical velocity of our high-mass He star models is small compared to previous models and decreases with increase of initial model mass, and the models parameters are shown in Table \ref{tab:1}.

\begin{table}[]
\caption{Initial HeZAMS parameters for rotating high-mass He star models with $\omega/\omega_{\text {crit }}$ =0.8}
\begin{threeparttable}
\begin{tabular}{ccccc}
\toprule
$M [M_{\odot}]$  & $T_{\mathrm{eff}} [\mathrm{kK}]$\tnote{1}  & log$L [L_{\odot}]$ & $R [R_{\odot}]$ & $V_{\mathrm{crit}} [\mathrm{km}/\mathrm{s}]$\tnote{2} \\
\midrule
35 & 72.4 & 6.00 & 6.40 & 242 \\
30 & 75.2 & 5.92 & 5.38 & 294 \\
25 & 82.6 & 5.80 & 3.92 & 442 \\
20 & 95.3 & 5.66 & 2.50 & 629 \\
\bottomrule
\end{tabular}
\begin{tablenotes}
\footnotesize
\item[1] Stellar effective temperatures \item[2] Critical rotational velocity at equator
\end{tablenotes}
\label{tab:1}
\end{threeparttable}
\end{table}
We take the 30 $M_\odot$ He stars with or without rotation as an example, as shown in the Figure \ref{fig:02r}.
The HR diagram shows that rotating He star in the early stages of evolution has higher effective temperatures, i.e. smaller radius. Because our mass-loss rate formula has a rotationally enhanced term, the rotating He star has a higher mass-loss rate, which reduces the radius of EIS.
As shown in the middle panel of Figure \ref{fig:02r}, the rotational velocity of high-mass He stars decreases to less than 100 $\mathrm{km}\cdot\mathrm{s}^{-1}$ in $\sim4 \times10^{4} \mathrm{yr}$ due to the high mass-loss rate that takes away angular momentum rapidly.
Since only one tenth of the lifetime has a rapid rotation, the rotational mixing increases the lifetime by $10^{4} \mathrm{yr}$ and also an additional loss of only 0.5 $M_{\odot}$ during the entire lifetime.
Since the rotation decay timescale is only one-tenth of the evolutionary lifetime, most observed WR stars are expected to be non-rapidly rotating and most of the rapid rotating WR stars are WN stars.
So far, 10 of the 70 observed samples analyzed of WR stars have line effects in their spectropolarimetry, except WR137, which is a WC7 star in a binary, and the other 9 are WN stars\citep{Vink2017b}.

As shown in the middle panel of Figure \ref{fig:02r}, the radiation zones at 0.76 $R_{\odot}$ to 0.96 $R_{\odot}$ near the convective convection core with $H_{\mathrm{p}}\sim10^{-0.8} R_{\odot}$ and $D_{\mathrm{rot}}\sim10^{7.0} \mathrm{cm}^2 \mathrm{s}^{-1}$, and the rotational mixing timescale $\tau\sim5\times10^{5} \mathrm{yr}$. Since the rotational mixing timescale $\tau\sim5\times10^{5} \mathrm{yr}$ is already larger than the lifetime $\sim4\times10^{5} \mathrm{yr}$ and rotational decay timescale $\sim4\times10^{4} \mathrm{yr}$ of He stars, rotational mixing has little effect on the evolution of high-mass He stars, which is consistent with the previous results in \cite{Limongi2018} who investigated the rotating massive stars during He-burning phase.

\subsection{H-free WR stars in the MW }
\begin{figure*}[htb]
\centering
\includegraphics[width=\textwidth]{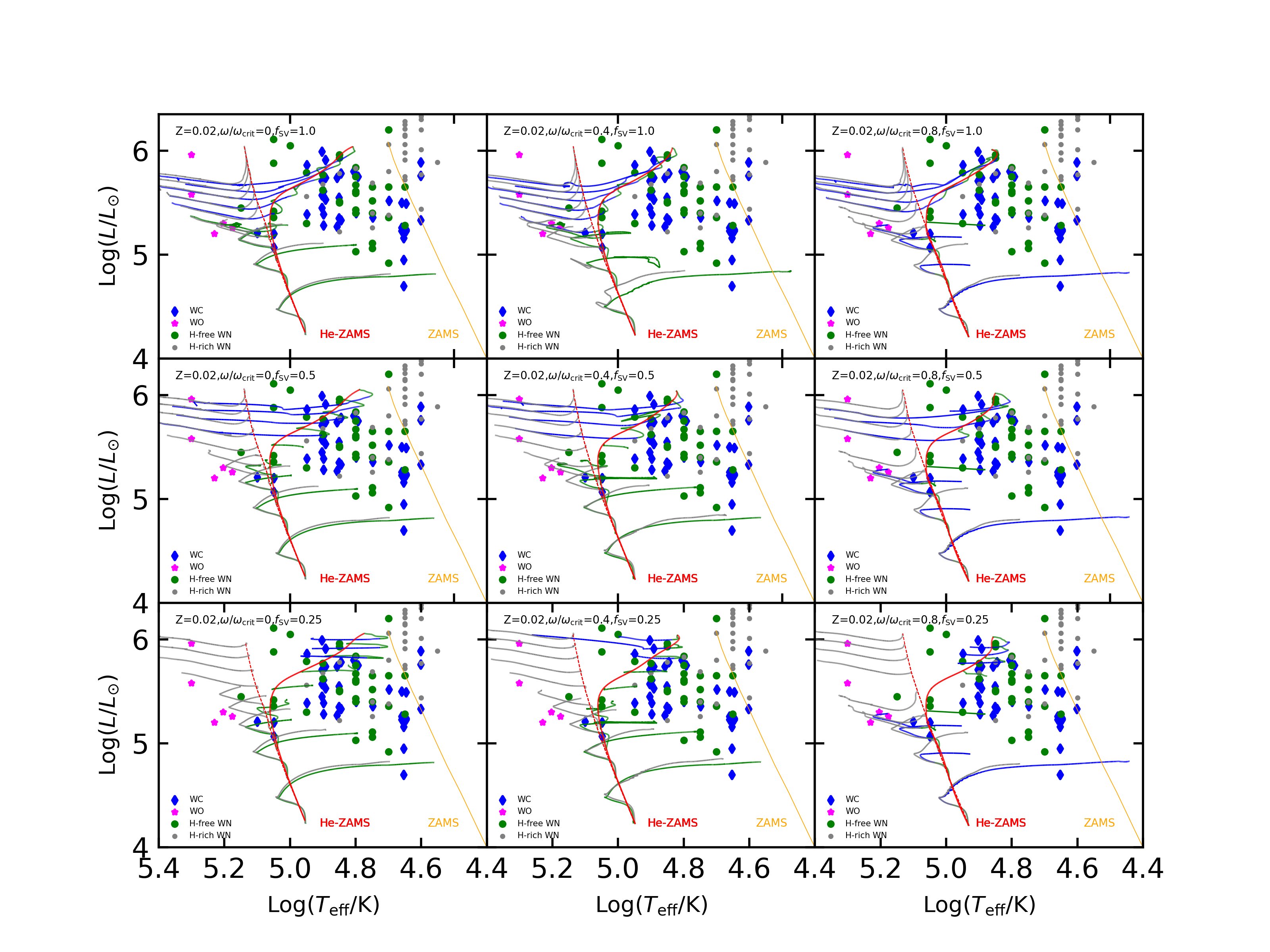}
\caption{Positions of H-free WR stars observed in the MW and evolutionary tracks of He stars with different masses. The initial masses from bottom to top are 4, 6, 8, 12, 16, 20, 25, 30, and 35 $M_{\odot}$. The input parameters are given in the top left corner of every panel.
The green and blue lines represent the WN and WC sequences, respectively. The gray lines represents the evolutionary tracks of the He stars without the EIS obtained using MLT++. The red dotted lines represent the HeZAMS without the EIS, while the red solid lines represent the HeZAMS with the EIS. The orange lines correspond to the ZAMS of normal stars. The observational data of the WN, WC, and WO stars in the MW were taken from \cite{Hamann2019}and \cite{Sander2019}.}
\label{fig:002}
\end{figure*}

In the MW, there are 667 Wolf-Rayet (WR) stars listed in the catalog \citep{Crowther2015}. The most recent data for 103 WR stars, with distance corrections based on Gaia DR2 data, have been reported by \cite{Hamann2019} and \cite{Sander2019}. These data are illustrated in Figure \ref{fig:002}.
The evolutionary tracks of He stars with different masses, rotational velocities and wind factors are also illustrated in Figure \ref{fig:002}. As discussed in the last subsection , the rotation and the wind factor have only limited effect on the evolutionary tracks. Although low-mass He stars with and without rapid rotation have similar evolutionary tracks, they have different chemical abundances on their surface because of the mixing triggered by the rapid rotation. Low-mass He stars subject to rapid rotation can explain the origin of WC stars with a low luminosity(log$(L/L_{{\odot}})\leq5.2$), while non-rotating low-mass He stars can evolve into H-free WN stars with low luminosity (log$(L/L_{{\odot}})\leq5.2$). Low luminosity (log$(L/L_{{\odot}})\leq5.2$) H-free WR stars can be understood in terms of the low-mass He stars in the He giant phase, while the WC stars among them require rapidly rotating low-mass He stars.

Although the EIS hardly affects the evolution of low-mass He stars, high-mass He stars with and without the EIS have very different evolutionary tracks. High-mass He stars without the EIS have a high effective temperature, and they can only explain the origin of WO stars. High-mass He stars with the EIS have a lower effective temperature and a larger radius. Their evolutionary tracks can explain the regions of log($T_{\rm eff}/$K)$\geq4.7$ and log$(L/L_{{\odot}})>5.7$. As shown in Figure \ref{fig:002}, high-mass He stars with a low wind factor ($f_{\mathrm{sv}}=0.25$) can evolve into H-free WN stars or WC stars but not WO stars (see the bottom panels), while those with high wind factor can become H-free WN, WC, or WO stars. However, our models still cannot accurately explain H-free WR stars with luminosities between $10^{5.2}$ and $10^{5.7} L_{{\odot}}$ and effective temperatures below $10^{4.9}$K nor H-free WR stars with luminosities above $10^{5.7} L_{{\odot}}$ and effective temperatures below $10^{4.7}$K. A possible reason is that our model underestimates the effect of the EIS on the stellar radius, that is, the EIS may work in the model of low-mass He stars or may result in larger stellar radii.

\begin{figure*}[htb]
\centering
\includegraphics[width=\textwidth]{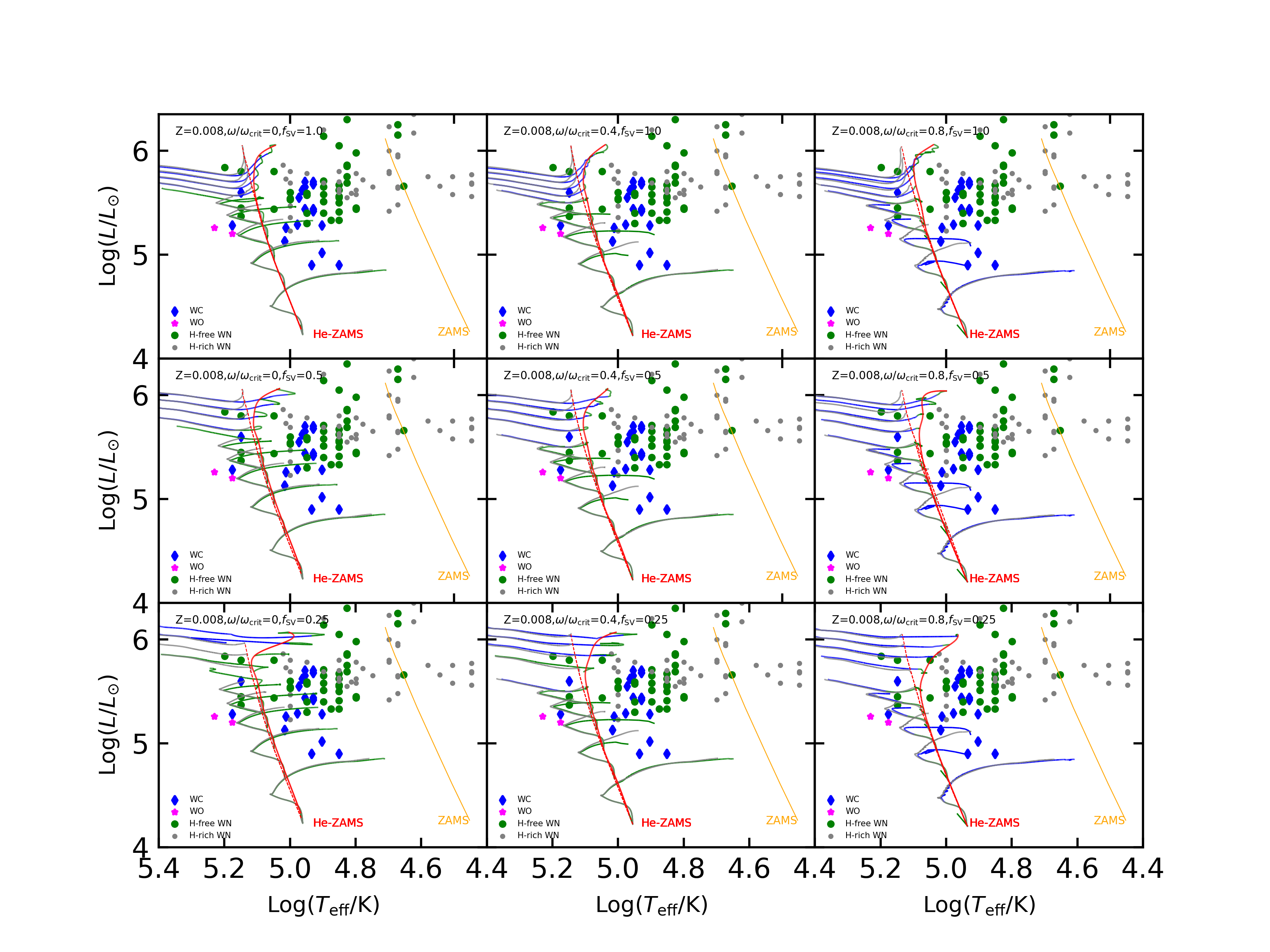}
\caption{Same as Fig.\ref{fig:002}, but for the models with Z=0.008. The observational data of the WN, WC, and WO stars in the LMC were taken from \cite{Crowther2002}, \cite{Hainich2014}, and \cite{Neugent2017}.}
\label{fig:008}
\end{figure*}

\subsection{H-free WR stars in the LMC}
In the LMC, there are 154 WR stars \citep{Neugent2018}. The 120 WR stars included in the catalog (\cite{Breysacher1999}) and analyzed (\cite{Crowther2002,Hainich2014}) are illustrated in Figure \ref{fig:008}. The evolutionary tracks of He stars with different masses, rotational velocities and wind factors are also shown in Figure \ref{fig:008}.
We note that the low luminosity(log$(L/L_{{\odot}})<5.2$) H-free WR stars in LMC are all WC stars, which means that the progenitors of these WC stars are subject to rapid rotation. As shown in Figure \ref{fig:008}, He stars with a rotation rate of $\omega / \omega_{c} = 0.8$ can evolve into WC stars in the helium giant phase and can be applied to the observed low luminosity WC stars.

Comparing Figures \ref{fig:002} and \ref{fig:008}, it can be observed that the effective temperature of H-free WR stars in the LMC is usually higher than that of H-free WR stars in the MW. Similarly, in our LMC models with $Z=0.008$, the evolutional tracks of high-mass He stars with the EIS shift toward a lower temperature to a lower extent. As shown in the lower left panel of Figure \ref{fig:002}, non-rotating high-mass helium stars with the EIS and small wind factors as well as and their evolutionary tracks can only explain the H-free WR stars on the right-hand side of the HeZAMS, but cannot explain most of the high luminosity(log$(L/L_{{\odot}})\geq5.2$) H-free WR stars.

To summarize, He stars with the EIS can evolve into some H-free WR stars in the galaxy, and very few H-free WR stars in the LMC. This indicates that the EIS may provide a strategy to understand the radius problem of H-free WR stars. At the same time, the one-dimensional models used in our work are not sufficiently detailed compared to the observed H-free WR stars. Several three-dimensional models have shown that the EIS is very complicated and differs from that of the one-dimensional models\citep{Moens2022}.
As the WR star wind is likely to be optically thick, the observed effective temperature may correspond to the temperature of the pseudo-photosphere\citep{Grassitelli2018, Sander2023}.
Therefore, more accurate wind models may also play a crucial role in solving the WR star radius problem\citep{Poniatowski2021, Lefever2023}.
In addition, H-free WR stars can also originate from the binary interaction, and they may be different from those simulated by our models\citep{Aguilera-Dena2022}. In future studies, we should consider the effects of binary interaction on H-free WR stars.

\section{Conclusions}
Using MESA, we simulate the evolution of He stars with and without MLT++ prescription. Due to the luminosity well below the Eddington limit, low-mass He stars with an initial mass of less than 12 $M_\odot$ do not produce the EIS with or without the MLT++ prescription.
High-mass He stars with an initial mass higher than 12 $M_\odot$ and without the MLT++ prescription produce the EIS, which significantly affects their structure and evolution.
The radiation luminosity in the FeCZ of high-mass He stars without MLT++ prescription is close to the Eddington limit.
The strong radiation pressure results in an extremely low density of the FeCZ, causing the radius of the FeCZ to increase and produce the EIS.
Their radius increases significantly from around 1-2 $R_\odot$ to around 5-10 $R_\odot$.

The metallicity, rotation and mass-loss rate affect the EIS.
Since EIS is $\Gamma$-dependent, its radius increases with the increase of metallicity and decreases with the increase of mass-loss rate.
The mass-loss rate increases with rotation, causing the radius of the EIS to decrease as rotation increases.
For the rotating low-mass He stars, the rotational mixing timescale ($2\sim4\times10^{5}\mathrm{yr}$) is smaller than their evolutionary timescale ($5\sim10\times10^{5}\mathrm{yr}$), and it can increase their core mass and allow these He stars to evolve into WC stars.
For the rotating high-mass He stars, the rotational mixing timescale ($4\sim10\times10^{5}\mathrm{yr}$) is larger than the rotational decay timescale ($3\sim9\times10^{4}\mathrm{yr}$) or even the lifetime ($3.5\sim5\times10^{5}\mathrm{yr}$). Rotation has little effect on the evolution of high-mass He stars.

The low luminosity (log$(L/L_{{\odot}})\leq5.2$) H-free WN stars in the MW and the LMC can be explained by the helium giant phase in low-mass He stars.
Low-luminosity WC stars with high surface $X_{\mathrm{C}}$ and $X_{\mathrm{O}}$ can be explained by rapidly rotating low-mass He stars.
High-mass He stars with the EIS can only explain H-free WR stars with a luminosity exceeding $10^{5.7} L_{{\odot}}$ and an effective temperature above $10^{4.7}$K in the MW. Due to the inefficient EIS caused by the low metallicity, high-mass He stars with the EIS can only explain H-free WR stars on the right-hand side of the HeZAMS in the LMC; most H-free WR stars with log $(L/ L_{{\odot}})\geq5.2$ in the LMC cannot be explained. High-mass stars with EIS evolve into WO stars only at the last evolution stage, and the lower lifetime fraction is consistent with the observed small number of WO stars. Those unexplained WR stars have larger radii and lower effective temperatures compared to our models. This discrepancy could be due to the pseudo-photospheres created by optically thick star winds or the need for our EIS models to take into account three-dimensional results or binary interaction.

\begin{acknowledgements}
We are grateful to anonymous referee for careful reading
of the paper and constructive criticism. This work received the generous support
of the National Natural Science Foundation of China, project Nos. U2031204,
12163005, 11863005, and 2022D01D85, the science research grants from the China
Manned Space Project with NO. CMS-CSST-2021-A10, and the Natural Science
Foundation of Xinjiang No. 2021D01C075.
\end{acknowledgements}

\bibpunct{(}{)}{;}{a}{}{,}
\bibliographystyle{aa}
\bibliography{aa}
\end{document}